# Polarization maintaining large mode area photonic crystal fiber

**J.R. Folkenberg[1], M.D. Nielsen[1,2], N.A. Mortensen[1], C. Jakobsen[1], and H.R. Simonsen[1]**

[1]*Crystal Fibre A/S, Blokken 84, DK-3460 Birkerød, Denmark*
[2]*COM, Technical university of Denmark, DK-2800 Kongens Lyngby, Denmark*

*jrf@crystal-fibre.com*

**Abstract:** We report on a polarization maintaining large mode area photonic crystal fiber. Unlike, previous work on polarization maintaining photonic crystal fibers, birefringence is introduced using stress applying parts. This has allowed us to realize fibers, which are both single mode at any wavelength and have a practically constant birefringence for any wavelength. The fibers presented in this work have mode field diameters from about 4 to 6.5 micron, and exhibit a typical birefringence of $1.5 \cdot 10^{-4}$.

## 1. Introduction

Since the invention of the so-called endlessly single-mode fiber [1], photonic crystal fibers (PCFs) have attracted much attention, because of the many novel fiber properties that may be realized. For a recent review see [2]. In particular, the ability to make large-mode area fibers

(LMA-fibers) which are single-mode at any wavelength and the large birefringence that may be achieved [3,4] is very interesting for fiber devices such as fiber lasers and fiber based gyroscopes.

LMA-fibers have been studied quite thoroughly [5], and it has been shown that the practical bandwidth of single-mode operation is limited at short wavelengths by micro- and macro bending losses, and at long wavelength by leakage losses. Different approaches to polarization maintaining fibers have also been investigated [3,4,6], all of which are based on form birefringence. Generally, the form birefringence is largest for wavelengths, $\lambda$, close to the pitch, $\Lambda$, of the cladding hole structure [4], but for larger values of $\Lambda/\lambda$ the birefringence decreases rapidly such that LMA-fibers with a birefringence on the order of $10^{-4}$ cannot be realized using form birefringence.

It is well known from solid silica fibers, that birefringence may be introduced using stress-applying parts (SAP), for an overview of conventional polarization maintaining (PM) fibers see [7]. That is, a part of the fiber consists of a material with a different thermal expansion coefficient than that of silica which gives rise to a stress field in the fiber when it is cooled below the softening temperature of silica during fabrication. In this way the fiber is given a built-in stress field and, because of the elasto-optic effect, the glass becomes birefringent.

In the present work we report on the first combination of PCF LMA-fibers and SAPs, to make polarization maintaining large-mode area fibers (PM-LMA).

## 2. Description of fibers

In Fig. 1, a microscope image of the realized PM-LMA structure is shown. It consists of an undoped silica core region surrounded by four periods of air holes in a triangular lattice, which forms the cladding. Outside the cladding, two SAPs are placed opposite to each other. The SAPs are made of boron-doped silica, with properties similar to the ones used for PANDA fibers [7].

The core and cladding region of the fiber resembles the so-called endlessly single-mode fibers [1], where the pitch, $\Lambda$, and hole size, d, has been chosen to minimize leakage and macro bending losses in the visible to near infrared region. Three different values of the pitch are realized, all having a relative holes size, $d/\Lambda$, of approximately 0.48. In the following the three fibers will be referred to as PM-LMA-1 ($\Lambda = 3.20$ μm), PM-LMA-2 ($\Lambda = 4.40$ μm) and PM-LMA-3 ($\Lambda = 5.94$ μm). The fibers have all been fabricated from the same preform, by varying the outer diameter of the fibers.

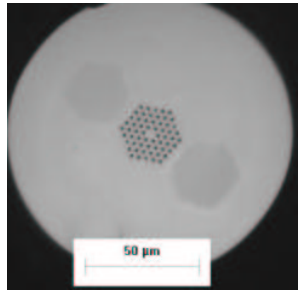

Fig. 1. Microscope image of PM-LMA fiber.

## 3. Optical properties

In Fig. 2, the spectral attenuation of the three fibers is shown, measured using the conventional cut-back method with a Tungsten-halogen white light source. For PM-LMA-1 an increased attenuation is observed at long wavelengths, which is ascribed to leakage losses [8]. The attenuation spectra indicate that the fiber is single-mode at all wavelengths, since no spectral features are observed which may be ascribed to a higher-order mode cutoff. This was verified by inspecting the near fields of the fibers at 635 nm and 1550 nm, showing only the fundamental mode.

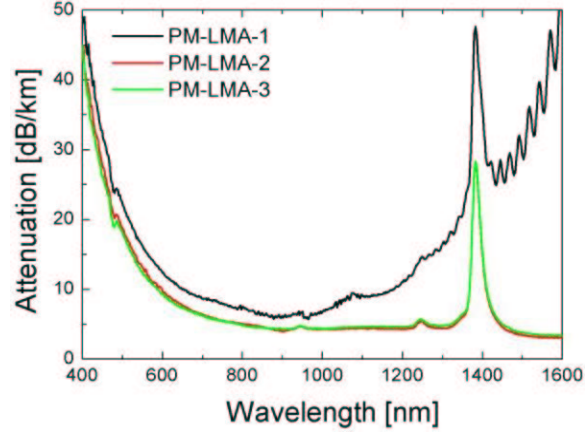

Fig. 2. Spectral attenuation for the three fibers.

The birefringence of the fibers was measured using the so-called crossed polarizer technique [9], illustrated in Fig 3. Linearly polarized white light is launched into the fiber with the polarization oriented at 45 degrees with respect to the polarization axis of the fiber. On the fiber output, a second polarizer (analyzer) is oriented at 45 degrees with respect to the polarization axis of the fiber. The transmitted light is focused into a multi-mode fiber and the spectrum recorded with an optical spectrum analyzer. Using a Jones matrix analysis of the setup, it may be shown that the intensity transmission is given by:

$$T = \cos^2(\Delta\beta L) \quad (1)$$

Here $\Delta\beta = \beta_x - \beta_y$ is the difference between the propagation constants of the polarization axes of the fiber and L is the length of the fiber. The crossed polarizer setup yields a transmission with spectral oscillations, the period of which may be related to the birefringence. In Fig. 4, typical spectra are shown for PM-LMA-1 at long wavelengths and PM-LMA-3 at short wavelengths.

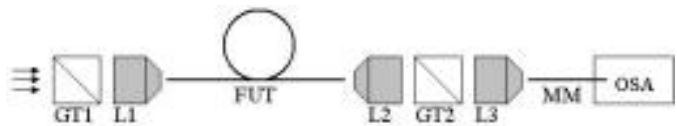

Fig. 3. Measurement set-up for group birefringence. GT1 : Input Glan-Thomson polarizer. L1 : Coupling lens into fiber under test (FUT). L2 : Collamating lens for fiber output. GT2 : Output Glan-Thomson polarizer (analyser). L3 : Focussing lens. MM : Multi-mode fiber for light collection and detection in optical spectrum analyser (OSA).

It is emphasized, that the analysis of the spectral oscillations yields the group birefringence, which may be shown by calculating the derivative of the phase in Eq. (1), $\varphi = \Delta\beta L$, with respect to the wavelength:

$$\frac{\partial \varphi}{\partial \lambda} = \frac{2\pi L}{\lambda^2} \left[ \lambda \frac{\partial \Delta n_{ph}}{\partial \lambda} - \Delta n_{ph} \right] = \frac{2\pi L}{\lambda^2} \Delta n_g \quad (2)$$

Here $\Delta n_{ph}$ is the phase birefringence and $\Delta n_g$ is the group birefringence. One oscillation period, $\Delta\lambda$, of the spectrum in Fig. 4 corresponds to $\Delta\varphi = 2\pi$, from which the group index on the right hand side of Eq. (2) may be calculated. The results are shown in Fig. 5, and have been averaged over intervals of 100 or 200 nm.

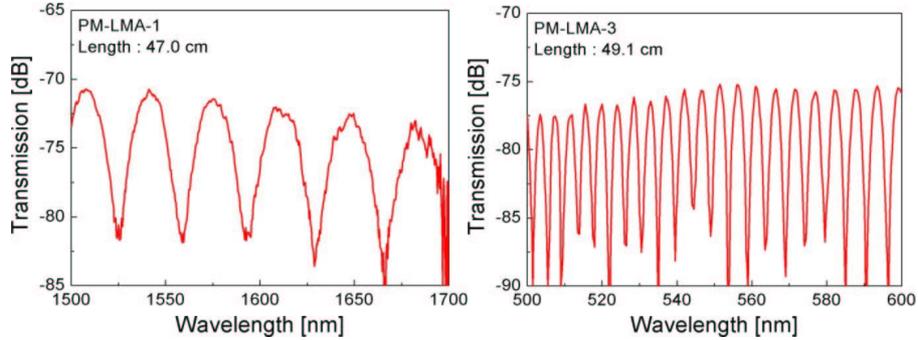

Fig. 4. Measured polarization osciallations, for the largest (left) and smallest (right) values of the relative wavelength, $\lambda/\Lambda$.

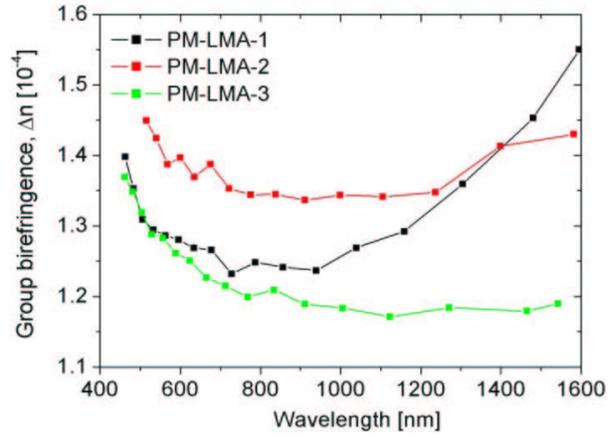

Fig. 5. Group birefringence vs. wavelength for the three fibers.

Finally, the polarization holding parameter (h-parameter) was also investigated using the setup shown in Fig. 3. For this experiment the first polarizer was aligned with one of the optical axis of the fiber, and at the output the transmitted intensity in the polarization axis was measured using the second polarizer. For PM-LMA-1 we obtained 13.3 dB extinction on a length of 390 m at 650 nm, corresponding to $h=1.2 \cdot 10^{-4}$ m$^{-1}$. This value may be influenced by many intrinsic or extrinsic factors, such as the fiber homogeneity, the wavelength, the bending radius and, probably most important, the magnitude of the birefringence itself. However, this investigation will not be discussed further here.

## 4. Discussion of results

The likely contributions to the birefringence in the present fiber may be divided into the categories known from solid-glass fibers, namely form birefringence and stress-induced birefringence. It has previously been shown that the six-fold symmetry of the cladding of large-mode area fiber does not allow inherent birefringence [10]. Furthermore, any asymmetry of the core region will only contribute negligibly to the birefringence measured, since the magnitude of the relative wavelength, λ/Λ, is too large [4]. Hence, the birefringence is expected to originate only from the stress field, and since the mode field diameter is known to be almost constant for LMA fibers [5], we may expect that the birefringence does not depend on the wavelength as it does for step-index fibers at small V-parameters [7].

Indeed, the values plotted in Fig. 5 show that the birefringence is practically constant for the three fibers that have been studied here. The small birefringence variation observed is attributed to diameter variations of the SAPs used in the preform, which are typically 10% along the length. Indeed, the SAP diameter as function of position in the preform correlates very well with the observed birefringence variations. Because the wavelength dependence is very small, we may assume that the term $\lambda \, d/d\lambda(\Delta n)$ in Eq. (2) is negligible and the phase birefringence is equal to the group birefringence.

For all three fibers the birefringence is increasing for decreasing wavelengths in the 400 nm – 900 nm region of the spectrum. We attribute this effect to the dispersion of the elasto-optic coefficient, which was discovered for bulk glasses in the beginning of the 20$^{th}$ century [11,12], but to our knowledge never reported for optical fibers. The birefringence is increasing in PM-LMA-1 for λ > 900 nm and to a smaller extend in PM-LMA-2 for λ > 1200 nm. In PM-LMA-1 this coincides with the spectral region showing leakage losses, and may be explained by the gradual expansion of the mode into the cladding region. Since the stress field cannot exist in the air holes, the magnitude of the stress field in the silica regions between the 6 air holes must be larger than in the core region itself, explaining the observed increase. For PM-LMA-2 no leakage losses are seen, but the increase of birefringence could be explained by dispersion of the elasto-optic coefficient at long wavelengths.

The SAPs used for the present fibers were designed to yield a birefringence of $3 \cdot 10^{-4}$ if used in a solid-glass PANDA fiber with similar SAP sizes and spacing. The reason for the lower birefringence in the present fibers is attributed to screening of the strain field by the holes. For d/Λ=1 the core is mechanically detached from the SAPs and thus not strained, while for d/Λ=0 the fiber is solid and thus comparable to a conventional PANDA fiber. Hence as d/Λ increases above zero, the strain experienced by the core is expected to decrease if the rest of the structure is kept constant. Yet, the level of birefringence obtained here is promising for fiber designs where SAPs are placed outside a cladding with air holes, and the result is somewhat surprising in the light of recent work. E.g. in [13] it is shown that an externally applied stress field on a photonic crystal fiber, obtained by a lateral force on the fiber cladding, yields a significant reduction of the birefringence even for relative hole sizes, d/Λ, smaller than 0.5. For the fibers studied here, the screening of the stress field may be reduced further, either by decreasing the number of periods of air holes or by using fiber designs with smaller holes [14].

## 5. Conclusion

We have demonstrated, to our knowledge for the first time, a polarization maintaining large mode area endlessly single mode PCF. The birefringence is induced by SAPs positioned outside the cladding region of the PCF. The optical properties of the LMA-PCF lead to a close to wavelength independent birefringence in the order of $1.5 \cdot 10^{-4}$.

**Acknowledgements**

M.D. Nielsen acknowledges financial support by the Danish Academy of Technical Sciences.